\documentclass[aip,jmp,amsmath,amssymb,preprint,]{revtex4-1}
\usepackage{color}
\usepackage{mathrsfs,amsmath}
\usepackage{mathrsfs}
\usepackage{graphicx}
\usepackage{dcolumn}
\usepackage{bm}
\usepackage{natbib}
\usepackage{amsthm}

\begin{document}
\newtheorem{thm}{Theorem}[section]
\newtheorem{lemma}[thm]{Lemma}
\newtheorem{prop}[thm]{Proposition}
\newtheorem{rem}[thm]{Remark}
\newtheorem{cor}[thm]{Corollary}

\title{\bf Thermodynamics of Markov Processes with Non-extensive Entropy and Free Energy}

\author{Liangrong Peng$^{1,2}$, Hong Qian$^3$, Liu Hong$^{2,3}$\\
$^1$College of Mathematics and Data Science, Minjiang University, Fuzhou, 350108, P.R.C.\\
$^2$Zhou Pei-Yuan Center for Applied Mathematics, Tsinghua University, Beijing, 100084, P.R.C.\\
$^3$Department of Applied Mathematics, University of Washington, Seattle, WA 98195-3925, U.S.A.
}
\homepage{Author to whom correspondence should be addressed. Electronic mail:  zcamhl@tsinghua.edu.cn, hqian@uw.edu}

\date{\today}

\begin{abstract}
Statistical thermodynamics of small systems shows dramatic differences from normal systems.  Parallel to the recently presented steady-state thermodynamic formalism for master equation and Fokker-Planck equation, we show that a ``thermodynamic'' theory can also be developed based on Tsallis' generalized entropy
$S^{(q)}=\sum_{i=1}^N(p_i-p_i^q)/[q(q-1)]$ and Shiino's generalized free energy
$F^{(q)}=[\sum_{i=1}^Np_i(p_i/\pi_i)^{q-1}-1]/[q(q-1)]$, where $\pi_i$ is
the stationary distribution.
$dF^{(q)}/dt=-f_d^{(q)}\le 0$ and it is zero iff the system is in its stationary
state. $dS^{(q)}/dt-Q_{ex}^{(q)} = f_d^{(q)}$ where $Q_{ex}^{(q)}$
characterizes the heat exchange.  For systems approaching equilibrium
with detailed balance, $f_d^{(q)}$ is the product of Onsager's thermodynamic
flux and force.  However, it is discovered that the Onsager's force
is non-local.  This is a consequence of
the particular transformation invariance for zero energy
of Tsallis' statistics.

\end{abstract}

\maketitle

\section{Introduction}

The Boltzmann-Gibbs (B-G) statistical thermodynamics is one of the fundamental theories of modern physics with wide applications in chemistry, biology, and even social sciences.
Recently, a complete statistical thermodynamic formalism based on both the B-G entropy and B-G free energy (relative entropy) has been presented for stochastic dynamics based on master equations \cite{ge2010,esposito2010} and Fokker-Planck equations \cite{Ge2009Extended,van2010}.
The theory introduces several key quantities: entropy production
rate, free energy dissipation rate, excess and house-keeping heat, etc.

Statistical thermodynamics of small systems requires the traditional thermodynamic quantities such as energy and entropy to be non-extensive \cite{Hill1963Thermodynamics}.
This extension of the classical B-G entropy provides a framework to study the systems with long-range interactions, multi-fractal structures, long-term memories, or systems whose interaction with macroscopic parts cannot be ignored \cite{Abe2001Nonextensive}.
It is therefore natural to ask whether the thermodynamic
formalism can be developed in parallel for non-extensive entropy
and free energy, such as the Tsallis entropy \cite{tsallis1988}, Renyi entropy \cite{Renyi1959,Renyi1970Probability} and many others \cite{Kaniadakis2001Nonlinear,Abe2001Nonextensive}.

Here we focus on the Tsallis entropy, which enjoys the properties of concavity, experimental robustness (Lesche stability), non-extensivity (as a generalization of the B-G entropy), and the finiteness of the rate of entropy production \cite{Boon2005Special, beck2009generalised}. This work shows that a similar structure does exist for Tsallis thermodynamics, which therefore would be of interest to a wide range applications, like defect turbulence \cite{daniels2004defect}, optical lattices \cite{lutz2003anomalous}, mathematical finance \cite{borland2002option}, and so on.

{\bf\em Generalized Entropy.}  The notions of Shannon's entropy
in the form $-\sum_{i=1}^N f_i\ln f_i$ and relative
entropy in the form $\sum_{i=1}^N f_i\ln(f_i/p_i)$ arise from
the asymptotic expressions for the probability of {\em
frequency} $\{f_i\}$ within a large set of independent and identically
distributed (i.i.d.) random samples, each with an equal probability
of $\frac{1}{N}$ and $\{p_i\}$ respectively.   The exponent
$I(f_1,\cdots,f_N)$ in the asymptotic expression $e^{-NI(f_1,\cdots,f_N)}$, known as large deviations rate function, provides the divergence function for an information geometry \cite{amari2007methods}. When a set of samples are not independent,
the probability for the frequency is no longer a simple
multi-nomial distribution; Tsallis entropy is believed to arise
in this context in connection with q-binomial distribution
\cite{kac2001quantum}.  The Shannon entropy
is intimately related to the logarithmic function, $y=\ln(x)$,
which is the inverse of exponential function $x=\exp(y)$.  The
exponential function is defined through the limit:	
\begin{equation}
	\exp(y) = \lim_{r\rightarrow 0}
	\left(1+ry\right)^{\frac{1}{r}}.
\label{def_e}
\end{equation}
The inverse function of the right-hand side of Eq. (\ref{def_e})
is
\begin{equation}
	y = \frac{x^r-1}{r}.
\label{eq_2}
\end{equation}
If one takes this as the analogue of the $y=\ln(x)$,
then corresponding to Gibbs' entropy $-\sum_i p_i\ln p_i$,
one can have a generalization of the entropy:
\begin{equation}
	-\sum_{i=1}^N p_i\left(\frac{p_i^r-1}{r}\right).
\label{t_s_e}
\end{equation}
If we denote $r=q-1$, then Eq. (\ref{t_s_e}) becomes
\begin{equation}
	S^{(q)} =   \sum_{i=1}^N p_i\bigg[\frac{1- p_i^{q-1}}{q(q-1)}\bigg],
\end{equation}
as $q\rightarrow1$. This is the non-extensive entropy first proposed by
Tsallis \cite{tsallis1988}, and $q\in \mathbb{R}$ ($q\neq 0, 1$) is known as the non-extensive parameter. Here we added an additional factor $q$ to the denominator to maintain the definite sign of free energy and its dissipation rate, which will be shown later. Because of relations in Eqs. (\ref{def_e}) and
(\ref{eq_2}), the Tsallis entropy enjoys many of the properties
of the Gibbs' entropy.

{\bf\em Relative entropy.} Recognizing the
correspondence between $\ln(x)$ and
$(x^{q-1}-1)/[q(q-1)]$ as $q\rightarrow1$, a generalization of the relative entropy \cite{hobson1969,Shore1980Axiomatic} with
respect to a distribution $g_i$ can be similarly
introduced \cite{shiino1998}:
\begin{equation}
	H^{(q)}(\{p_i\}||\{g_i\}) = \frac{1}{q(q-1)}
		\left[\sum_{i=1}^N p_i \left(\frac{p_i}{g_i}
			\right)^{q-1}-1\right].
\end{equation}

{\bf\em Statistical internal energy and \boldmath{$H^q$} as the
Gibbs free energy.}
Assuming $\{\pi_i\}$ is a stationary distribution to the master
equation in Eq. (\ref{master_eq}) below, one can define a statistical
internal energy by applying the Boltzmann's law in reverse \cite{ge2010}:
\begin{equation}
	\epsilon_i^{(q)} = \frac{\pi_i^{1-q}-1}{q(q-1)},
\label{energy}
\end{equation}
in analogous to B-G energy $\epsilon_i=\ln(1/\pi_i)$.
Then the mean statistical internal energy is
\begin{equation}
	U^{(q)} = \sum_{i=1} p_i^q\epsilon_i^{(q)}
	= \sum_{i=1}^N p_i^q \left[ \frac{\pi_i^{1-q}-1}{q(q-1)} \right].
\end{equation}
The mean internal energy being the sum of $p_i^q\epsilon_i^{(q)}$ rather
than the conventional $p_i\epsilon_i$ is a peculiar feature of the
Tsallis statistics.  Then we have generalized free energy:
\begin{equation}
	F^{(q)} = U^{(q)}-S^{(q)} = \frac{1}{q(q-1)}\left[\sum_{i=1}^Np_i\left(\frac{p_i}
		{\pi_i}\right)^{q-1} - 1\right]
			= H^{(q)}(\{p_i\}||\{\pi_i\}).
\end{equation}

{\bf\em Conditional entropy.} Tsallis entropy
does not enjoy the standard equation concerning
conditional entropy of a pair of correlated random
variables $A$ and $B$ \cite{Khinchin1953The,Shore1980Axiomatic}:
\begin{subequations}
\label{eq_9}
\begin{equation}
	    S(AB) = S(A)+S_A(B),
\end{equation}
where
\begin{eqnarray}
	S(AB) &=& -\sum_{i,j}p_iq_{ij}\log\left(p_iq_{ij}\right),
\\
	S(A) &=& -\sum_{i}p_i\log p_i,
\\
	S_A(B) &=& -\sum_{i}p_i\left(
		\sum_jq_{ij}\log q_{ij} \right).
\end{eqnarray}
\end{subequations}

For Tsallis's entropy, one has
\begin{eqnarray*}
	S^{(q)}(AB) &=& \frac{1}{q(q-1)}\left(1-\sum_{i,j} (p_iq_{ij})^q
		\right)= \frac{1}{q(q-1)}\left[1- \sum_i p_i^q
			+\sum_i p_i^q -\sum_{i,j} (p_iq_{ij})^q
		\right]
\\
	&=& S^{(q)}(A) + \sum_i p_i^q \left[
			\frac{1 -\sum_j q_{ij}^q}{q(q-1)} \right]= S^{(q)}(A) + \sum_i p_i^q\  S^{(q)}(B|A=i)
\\
	&\le&  S^{(q)}(A) + \sum_i p_i\ S^{(q)}(B|A=i)= S^{(q)}(A)+S_A^{(q)}(B),
\end{eqnarray*}
when $q>1$. And the inequality sign changes when $q<1$.

Eq. (\ref{eq_9}), together with
{\em entropy reaches its maximum with uniform distribution}
\begin{equation}
	S\left(\{p_1,p_2,p_3,\cdots,p_N\}\right)
		\le S\left(\left\{\frac{1}{N},\frac{1}{N},\frac{1}{N},\cdots,							\frac{1}{N}\right\}\right),
\label{eq_10}
\end{equation}
and {\em entropy is invariant with adding an impossible event}
\begin{equation}
	S\left(\{p_1,p_2,p_3,\cdots,p_N,0\}\right)
		= S\left(\{p_1,p_2,p_3,\cdots,p_N\}\right),
\label{eq_11}
\end{equation}
providing Khinchin's uniqueness theorem for Gibbs entropy \cite{Khinchin1953The}. It is noted that Tsallis entropy does satisfy both Eqs. (\ref{eq_10}) and (\ref{eq_11}). By replacing Eq. (\ref{eq_9}) with a more general non-extensive one, Abe has shown that the Tsallis entropy follows uniquely from the generalized version of Khinchin's axioms \cite{abe2000axioms}.

This paper is organised as follows. In Sec. 2 and 3, the Tsallis thermodynamics of master equations and F-P equations are constructed separately, with emphasis on the correct non-negative form of house-keeping heat. In Sec. 4, the underlying mathematical connection and correspondence between the two formalisms is explored and justified for both the discretized F-P equations and tridiagonal master equations, in the limit of infinitesimal jump step in space. Sec. 5 provides two simple but insightful examples to illustrate the application of Tsallis thermodynamics and its difference from B-G thermodynamics. Finally, we summarize our results in Sec. 6.  Proofs on theorems in Sec. 4 can be found in appendix for completeness.

\section{Tsallis Thermodynamics of Master Equations}
\label{Tsallis Thermodynamics of Master Equations}


We now consider a Markov process with master equation
\begin{equation}
	\frac{dp_i(t)}{dt} = \sum_{j=1}^N \left(p_jw_{ji}-p_iw_{ij}\right), \quad i=1,2,\cdots,N,
\label{master_eq}
\end{equation}
which has been widely applied for modeling various irreversible processes, including quantum physics, chemical reactions and open biological systems \cite{van1992stochastic,Tanimura2006Stochastic,Ge2016Mesoscopic}.

{\bf\em Relative entropy and positive free energy
dissipation.}
We assume both $\{p_i(t)\}$ and $\{g_i(t)\}$ satisfy the same master
equation, with different initial distributions.
Then the time derivative of the relative entropy
\begin{eqnarray}
	\frac{d}{dt}H^{(q)}(t) &=& \frac{1}{q(q-1)}\sum_{i=1}^N
		\left[ q\left(\frac{p_i(t)}{g_i(t)}\right)^{q-1}\frac{dp_i}{dt}
		-(q-1)\left(\frac{p_i(t)}{g_i(t)}\right)^{q}\frac{dg_i}{dt}\right]
\nonumber\\
	&=& \frac{1}{q(q-1)}\sum_{i,j=1}^N w_{ji}
		\left\{ qp_j\left[\left(\frac{p_i}{g_i}\right)^{q-1}
			-\left(\frac{p_j}{g_j}\right)^{q-1}\right]\right. \left.-(q-1)g_j\left[\left(\frac{p_i}{g_i}\right)^q -
			\left(\frac{p_j}{g_j}\right)^q\right]\right\}
\nonumber\\
	&=& \frac{1}{q(q-1)}\sum_{i,j=1}^N \frac{w_{ji}p_j^q}{g_j^{q-1}}
		\left\{ q\left(\frac{p_ig_j}{g_ip_j}\right)^{q-1}
			-(q-1)\left(\frac{p_ig_j}{g_ip_j}\right)^q -1
			\right\}\le 0.
\end{eqnarray}
The last step is because $w_{ij}\ge 0$ when $i\neq j$, the summand
is zero when $i=j$, and the term in the $\{\cdots\}$:
\begin{equation}\label{inequality}
	  \frac{qz^{q-1}-(q-1)z^q -1}{q(q-1)} = -\frac{(1-z)^2}{q(q-1)}
		\frac{d}{dz}\left(\frac{1-z^q}{1-z}\right) \le 0, \ \ \
		(z>0).
\end{equation}

	Now if we take a stationary distribution $\{\pi_i\}$ as
the $\{g_i\}$, then we have the time derivative of
generalized free energy $F^{(q)}$:
\begin{eqnarray}
	\frac{dF^{(q)}}{dt} &=& \frac{1}{q-1}\sum_{i,j=1}^N w_{ji}
		p_j\left[\left(\frac{p_i}{\pi_i}\right)^{q-1}
			-\left(\frac{p_j}{\pi_j}\right)^{q-1}\right]= -\frac{1}{2}\sum_{i,j=1}^N \phi_{ij}\Phi_{ij}^{(q)},
\end{eqnarray}
where
\begin{equation}
	\phi_{ij}(t) = p_iw_{ij}-p_jw_{ji},  \  \  \
	\Phi_{ij}^{(q)}(t) = \frac{\left(p_i/\pi_i\right)^{q-1}
			-\left(p_j/\pi_j\right)^{q-1}}{q-1}.
\end{equation}

{\bf\em Markovian dynamics with detailed balance.}
Master equation with detailed balance is a special
class of stochastic systems which represents closed systems
approaching to equilibrium.  For master equations with detailed
balance, $\pi_iw_{ij}=\pi_jw_{ji}$ is equivalent to the
Kolmogorov cycle condition:
\begin{equation}
	\frac{w_{i_0i_1}w_{i_1i_2}\cdots w_{i_{n-1}i_n}w_{i_ni_0}}
	{w_{i_1i_0}w_{i_2i_1}\cdots w_{i_ni_{n-1}}w_{i_0i_n}} = 1,
\end{equation}
for any cycle path $i_0,i_1,\cdots,i_{n-1},i_n,i_0$ \cite{Jiang2005Mathematical}.
Then we have
\begin{equation}
	\Phi_{ij}^{(q)}(t) = \left(\frac{2}{(\pi_jw_{ji})^{q-1}+
		(\pi_iw_{ij})^{q-1}}
			\right)
		\frac{\left(p_iw_{ij}\right)^{q-1}
			-\left(p_jw_{ji}\right)^{q-1}}{q-1}.
\label{O_force}
\end{equation}
In this case, the $\phi_{ij}$ and $\Phi^{(q)}_{ij}$ can be
identified as Onsager's thermodynamic flux and force,
respectively.  Therefore, we shall call their
product the {\em free energy dissipation rate}
\begin{equation} \label{master f_d}
	-\frac{dF^{(q)}}{dt} = f_d^{(q)} = \frac{1}{2}\sum_{i,j=1}^N \phi_{ij}\Phi_{ij}^{(q)}
		\ge 0.
\end{equation}

	For standard extensive thermodynamics, the Onsager's thermodynamic
force $\Phi_{ij}$ is ``locally determined''.  It depends only on
$p_i$, $p_j$, $w_{ij}$, $w_{ji}$ and the local energy
difference $\epsilon_i-\epsilon_j =\ln(w_{ij}/w_{ji})$.   However,
Eq. (\ref{O_force}) shows that the $\Phi_{ij}^{(q)}$ here
depends also on stationary $\pi_i$ and $\pi_j$, not just
their ratio which can be determined locally (Or equivalently,
it depends on the absolute value of $\epsilon_i^{(q)}$ and $\epsilon_j^{(q)}$,
not just their difference). This is a consequence of
the particular transformation invariance for zero energy
of Tsallis' statistics.

{\bf\em Entropy production rate.}
With or without detailed balance, we have
\begin{eqnarray}
	\frac{dS^{(q)}(t)}{dt} &=& -\frac{1}{q-1}
	\sum_{i,j=1}^N p_i^{q-1}\left(p_jw_{ji}-p_iw_{ij}\right)=\frac{1}{2}
	\sum_{i,j=1}^N \phi_{ij}\left(\frac{p_i^{q-1}-p_j^{q-1}}{q-1}
		\right)
\nonumber\\
	&=&  f_d^{(q)} - \frac{q}{2}\sum_{i,j=1}^N \phi_{ij}
		\left(p_i^{q-1}\epsilon_i^{(q)} - p_j^{q-1}\epsilon_j^{(q)}
			\right),
\label{de_comp}
\end{eqnarray}
where statistical internal energy $\epsilon_i^{(q)}$ is defined in
Eq. (\ref{energy}) from the stationary distribution $\pi_i$.
The last term in Eq. (\ref{de_comp}) can be identified as
{\em excess heat exchange rate} \cite{oono1998}:
\begin{equation}
		Q_{ex}^{(q)} = -\frac{q}{2}\sum_{i,j=1}^N \phi_{ij}
		\left(p_i^{q-1}\epsilon_i^{(q)} - p_j^{q-1}\epsilon_j^{(q)}
			\right).
\label{qex}
\end{equation}

The $Q_{ex}^{(q)}=0$ in the steady state.  However, for
equilibrium steady state with detailed balance, each term in
the summand is zero.  In contrast, for nonequilibrium steady
state, the summand is not zero. But the sum is zero due
to Tellegen theorem: The ``heat, or kinetic energy'' continuously
circulates in the system in nonequilibrium steady state.

Similar as in the B-G thermodynamics \cite{oono1998,ge2010,esposito2010},
the time derivative of Tsallis entropy can also be separated into two distinct parts
\begin{equation*}
	\frac{dS^{(q)}(t)}{dt}
	=  \frac{1}{2}
	\sum_{i,j=1}^N \phi_{ij}\left(\frac{p_i^{q-1}-p_j^{q-1}}{q-1}
		\right)
    \equiv e_p^{(q)} - h_d^{(q)},
\end{equation*}
where the {\em heat dissipation rate }
\begin{eqnarray}
	h_d^{(q)} &=& \frac{1}{2(q-1)}\sum_{i,j=1}^N \phi_{ij}\left[p_i^{q-1}\left(w_{ij}^{q-1}-1\right)
			-p_j^{q-1}\left(w_{ji}^{q-1}-1\right)\right],
\end{eqnarray}
and the {\em entropy production rate}
\begin{align}
e_p^{(q)}&=\frac{1}{q-1}\sum_{i,j}p_i w_{ij} \big[ (p_i w_{ij})^{q-1} - (p_j w_{ji})^{q-1}\big] \label{master e_p}
\\
&=\frac{1}{2(q-1)}\sum_{i,j} \phi_{ij}\big[ (p_i w_{ij})^{q-1} - (p_j w_{ji})^{q-1}\big]\nonumber
\geq 0.
\end{align}
The $e_p^{(q)}$ is greater than zero since if $p_iw_{ij}\ge p_jw_{ji}$,
then $\left((p_iw_{ij})^{q-1}-(p_jw_{ji})^{q-1}\right)/(q-1)\ge 0$.

{\bf\em Correct form of house-keeping heat.}
In B-G thermodynamics, the difference of entropy production rate and free energy dissipation rate $e_p^{(q)}-f_d^{(q)}$ is also known as the house-keeping heat, which is always greater than zero. However, in Tsallis thermodynamics, the sign of
\begin{align}
\widetilde{Q_{hk}^{(q)}}&=e_p^{(q)} - f_d^{(q)}=\frac{1}{q-1}\sum_{i,j}p_i w_{ij} \big[ (p_i w_{ij})^{q-1} - (p_j w_{ji})^{q-1} -  (\frac{p_i}{\pi_i})^{q-1} + (\frac{p_j}{\pi_j})^{q-1}\big],
\end{align}
is undefined. Consider a three-state Markov jump process and choose the transition rate matrix as
\begin{equation}
W=
\begin{pmatrix}
-1 & 1/2 & 1/2\\
1/2 & -1 & 1/2\\
1/2 & 1/2 & -1
\end{pmatrix},\nonumber
\end{equation}
whose equilibrium distribution is $\pi= (1/3,1/3,1/3)$.
Given the initial distribution $p(0)=(1, 0, 0)$ and $q=2$,
it is direct to see $\widetilde{Q_{hk}^{(q)}}(t=0)= -2.5<0$.

Therefore, it is natural to ask whether a part of $\widetilde{Q_{hk}^{(q)}}$ can play the same role of house-keeping heat while still being non-negative? Before introducing the correct formula, we show that, a direct generalization of house-keeping heat $\overline{ Q_{hk}^{(q)} }=\frac{1}{q-1}\sum_{i,j} p_i w_{ij}
\big[ ({\pi_i w_{ij}})^{q-1} -  ({\pi_j w_{ji}})^{q-1} \big]$ from B-G to Tsallis thermodynamics fails to do this.
In fact, even though $\lim_{q \rightarrow 1} \overline {Q_{hk}^{(q)}} = \sum_{i,j} p_i w_{ij} [\ln (\pi_i w_{ij}) - \ln (\pi_j w_{ji})]$ recovers the house-keeping heat in B-G thermodynamics, it is straightforward to see that by choosing $q=2$, the transition rate matrix
\begin{equation} \label{wness23}
W=
\begin{pmatrix}
-3 & 1  & 2 \\
2  & -3 & 1 \\
2 & 1  & -3
\end{pmatrix},\nonumber
\end{equation}
and the initial distribution $p(0)= (1 ,0, 0)$, we have $\overline{ Q_{hk}^{(q)} }(t=0) <0$.

To be correct, the difference between entropy production rate $e_p^{(q)}$ and free energy dissipation rate $f_d^{(q)}$ is decomposed into two parts: the house-keeping heat ${Q}_{hk}^{(q)} $ and non-extensive heat $Q_{ne}^{(q)}$, \textit{i.e.}
\begin{align}
&Q_{hk}^{(q)}
=\frac{1}{q-1}\sum_{i,j} p_i w_{ij}
\big[ ({\pi_i w_{ij}})^{q-1} -  ({\pi_j w_{ji}})^{q-1} \big]
- \frac{1}{q}\sum_{i,j} \frac{p_i}{\pi_i} \big[ ({\pi_i w_{ij}})^{q} -  ({\pi_j w_{ji}})^{q} \big] , \label{master Q_hk}\\
&Q_{ne}^{(q)}=\frac{1}{q-1}\sum_{i,j}p_i w_{ij}  \bigg\{ (p_i w_{ij})^{q-1} - (p_j w_{ji})^{q-1} -  (\frac{p_i}{\pi_i})^{q-1} + (\frac{p_j}{\pi_j})^{q-1} \nonumber \\
&~~~~~~~ - \big[ ({\pi_i w_{ij}})^{q-1} -  ({\pi_j w_{ji}})^{q-1} \big]
+ \frac{q-1}{q } \frac{1}{\pi_i w_{ij}} \big[ ({\pi_i w_{ij}})^{q} -  ({\pi_j w_{ji}})^{q} \big]
\bigg\} ,
\end{align}
where ${Q}_{hk}^{(q)}$ represents the distance from detailed balance condition, and $Q_{ne}^{(q)}$ arises due to the non-extensibility of Tsallis thermodynamics, which vanishes in the limit of $q\rightarrow 1$.
Based on the polynomial inequality, the sign of ${Q}_{hk}^{(q)}$ is readily determined as follows.

\begin{prop}
Let $\{\pi_i>0 \}_{i=1}^N$ be the steady state of master equations in \eqref{master_eq}, then the house-keeping heat
\begin{equation}
{Q}_{hk}^{(q)}
=\frac{1}{q-1}\sum_{i,j} p_i w_{ij}
\big[ ({\pi_i w_{ij}})^{q-1} -  ({\pi_j w_{ji}})^{q-1} \big]
- \frac{1}{q}\sum_{i,j} \frac{p_i}{\pi_i} \big[ ({\pi_i w_{ij}})^{q} -  ({\pi_j w_{ji}})^{q} \big]
\geq 0\nonumber,
\end{equation}
is non-negative definite, which becomes zero if and only if at the equilibrium state.
\end{prop}
\begin{proof}
The house-keeping heat for master equations can be reformulated as
\begin{align*}
 {Q}_{hk}^{(q)}
=&\frac{1}{q (q-1) }\sum_{i,j} p_i w_{ij} (\pi_i w_{ij})^{q-1}
\bigg[1 -  q (\frac{\pi_j w_{ji}}{\pi_i w_{ij}})^{q-1}  + (q-1) (\frac{\pi_j w_{ji}}{\pi_i w_{ij}})^{q}  \bigg]
 \geq 0,
\end{align*}
where in the last step we have used the inequality $\frac{1-qz^{q-1} + (q-1)z^q}{q(q-1)} \geq 0$ for $z=\frac{\pi_j w_{ji}}{\pi_i w_{ij}} >0$ as proved in (\ref{inequality}).
The $Q_{hk}^{(q)} =0 $ if and only if $z=\frac{\pi_j w_{ji}}{\pi_i w_{ij}} =1$ for each pair of $(i,j)$, that is, the system is at equilibrium.
\end{proof}

{\bf\em Laws of Tsallis thermodynamics.}
With above thermodynamic quantities in hand, laws of Tsallis thermodynamics for master equations are reformulated as
\begin{subequations}
\begin{eqnarray}
&&\frac{dS^{(q)}}{dt}+h_{d}^{(q)} =e_p^{(q)} \geq 0, \\
&&\frac{dS^{(q)}}{dt}- Q_{ex}^{(q)} =f_d^{(q)} \geq 0 ,\\
&&Q_{ex}^{(q)}+h_d^{(q)} - Q_{ne}^{(q)}= Q_{hk}^{(q)} \geq 0.
\end{eqnarray}
\end{subequations}
The first relation shows that the entropy production rate $e_p^{(q)}$ is always non-negative, as a manifestation of the second law of thermodynamics, while the other two relations go beyond the classical ones.

{\bf\em Connections with B-G thermodynamics.}
By recalling the correspondence between $q$-logarithm $(x^{q-1} - 1)/(q-1)$ and natural logarithm $\ln (x)$, as ${q \rightarrow 1}$, we have the entropy, internal energy and free energy in Tsallis thermodynamics become the corresponding items in B-G thermodynamics respectively \cite{ge2010,esposito2010}, \textit{i.e.}
\begin{subequations}
\begin{eqnarray}
&&\lim_{q \rightarrow 1} S^{(q)}=-\sum_{i=1}^N p_i \ln p_i,\\
&&\lim_{q \rightarrow 1} U^{(q)}=-\sum_{i=1}^N p_i \ln \pi_i,\\
&&\lim_{q \rightarrow 1}F^{(q)}=\sum_{i=1}^N p_i \ln (p_i/\pi_i).
\end{eqnarray}
\end{subequations}

Similar conclusions also hold for the entropy production rate, free energy dissipation rate and house-keeping heat:
\begin{align}
& \lim_{q \rightarrow 1} e_p^{(q)}=
\frac{1}{2}\sum_{i,j=1}^N \left( p_i w_{ij} - p_j w_{ji}\right) \ln \frac{p_i w_{ij}}{p_j w_{ji}}
\geq 0, \label{master ep rate boltz}\\
&\lim_{q \rightarrow 1} f_d^{(q)} =
\frac{1}{2}\sum_{i,j=1}^N \left( p_i w_{ij} - p_j w_{ji}\right) \ln \frac{p_i \pi_{j}}{p_j \pi_{i}} \geq 0, \label{master fd rate boltz}\\
&\lim_{q \rightarrow 1}Q_{hk}^{(q)}=
\frac{1}{2}\sum_{i,j=1}^N \left( p_i w_{ij} - p_j w_{ji}\right) \ln \frac{\pi_i w_{ij}}{\pi_j w_{ji}}
\geq 0
 \label{master Qhk rate boltz},
\end{align}
where the right-hand side of Eqs. \eqref{master ep rate boltz}-\eqref{master Qhk rate boltz} are identified as the corresponding entropy production rate, free energy dissipation rate and house-keeping heat in B-G thermodynamics of master equations \cite{ge2010,esposito2010}. Furthermore, limits of heat dissipation rate and change rate of excess heat
\begin{align}
& \lim_{q \rightarrow 1} h_d^{(q)}=
\frac{1}{2}\sum_{i,j=1}^N \left( p_i w_{ij} - p_j w_{ji}\right) \ln \frac{w_{ij}}{w_{ji}}
,\\
&\lim_{q \rightarrow 1}Q_{ex}^{(q)}=
\frac{1}{2}\sum_{i,j=1}^N \left( p_i w_{ij} - p_j w_{ji}\right) \ln \frac{\pi_i}{\pi_j},
\end{align}
agree exactly with the definitions in B-G thermodynamics.

It is noticeable that, the non-extensive heat reduces to zero as $q$ approaches to $1$, that is, $\lim_{q \rightarrow 1}Q_{ne}^{(q)}=0$. Thus the house-keeping heat equals to the difference between entropy production and free energy dissipation rates in B-G thermodynamics. This implies that $Q_{ne}^{(q)}$ can serve as a characteristic quantity to describe the non-extensivity of the system.

\section{Tsallis Thermodynamics of Fokker-Planck Equations}
\label{Tsallis Thermodynamics of Fokker-Planck Equations}

In this Section, we turn to the Tsallis thermodynamics of Fokker-Planck (F-P) equations, which read
\begin{equation}\label{FP}
\frac{\partial}{\partial t}  p(x,t) =  -\frac{\partial}{\partial x} K(x,t), \quad
K(x,t)=u(x) p - D(x)\frac{\partial p}{\partial x},
\end{equation}
where $u(x)$ and $D(x)>0$ denote the drift and diffusion coefficients respectively. In particular, the steady-state probability distribution is given by a time-independent solution of $\frac{\partial}{\partial t}  p(x,t)=0$, that is,
\begin{equation}
\frac{\partial}{\partial x} K^s(x) = 0, \quad
K^s(x)=u(x)\pi - D(x)\frac{\partial \pi}{\partial x},
\end{equation}
where $\pi(x)\geq0$ for $x\in \mathbb{R}$. Compared to master equations, the information of transition rate matrix is largely compressed to the draft and diffusion coefficients, which as a consequence will simplify the modeling procedure and numerical computations a lot.


{\bf\em Tsallis thermodynamics of F-P equations.}\label{TsallisFP}
By extending the state variables into continuous cases, we can directly borrow the definitions of the generalized entropy, internal energy and free energy from master equations and apply to F-P equations\cite{shiino1998},
\begin{subequations}
\begin{eqnarray}
&&S^{(q)}(t)=-\int dx p \left[\frac{p^{q-1}-1}{q(q-1)}\right], \\
&&U^{(q)}(t)=\int dx p^q\left[\frac{\pi^{1-q} -1}{q(q-1)}\right], \\
&&F^{(q)}(t)= \frac{1}{q(q-1)} \int dx \left[ p \left(\frac{p}{\pi}\right)^{q-1} - p\right],
\label{FP free energy}
\end{eqnarray}
\end{subequations}
where $(q\neq 0,1)$. The free energy dissipation rate $f_d^{(q)} = -{d F^{(q)}}/{dt}$ associated with F-P equations is given by
\begin{equation}\label{FP f_d}
\begin{split}
f_d^{(q)} &= - \frac{1}{q-1} \int {dx} \bigg[ K \frac{\partial}{\partial x}(\frac{p}{\pi})^{q-1} \bigg]
     = \frac{1}{q(q-1)} \int dx \bigg[q (\frac{p}{\pi})^{q-1} \frac{\partial K}{\partial x}
     -(q-1) (\frac{p}{\pi})^{q} \frac{\partial K^s}{\partial x} \bigg]  \\
    &=- \int dx \bigg[ K \frac{1}{q-1} \frac{\partial }{\partial x}(\frac{p}{\pi})^{q-1}
     -K^s  \frac{1}{q} \frac{\partial }{\partial x}(\frac{p}{\pi})^q \bigg]  = \int dx \bigg[ D \frac{\pi^2}{p}(\frac{p}{\pi})^{q-1} (\frac{\partial }{\partial x}(\frac{p}{\pi}))^2
    \bigg]\\
    &
    = \int dx \bigg[  \frac{p}{D}(\frac{p}{\pi})^{q-1} (\frac{ K }{p} - \frac{K^s}{\pi})^2
    \bigg] \geq 0,
\end{split}
\end{equation}
where in the second line we have used the integration by parts, and in the last line we have used the relation $(\frac{K}{p} - \frac{K^s}{\pi }) = -D \frac{\partial }{\partial x} \ln(\frac{p}{\pi})$.
As a result, the free energy dissipation rate is always non-negative \cite{shiino1998}, and vanishes if and only if in the steady state.
Notice that $f_d^{(q)}$ possesses a similar form as in classical B-G thermodynamics except for the factor $(\frac{p}{\pi})^{q-1}$.

Now we proceed to derive the balance equation of Tsallis entropy.
\begin{align*}
\frac{dS^{(q)}(t)}{dt}
&= -\int dx K p^{q-2}\partial_x p
=\int dx \bigg[  \frac{p}{D} p^{q-1} (\frac{K}{p})^2  - \frac{u}{D} p^{q-1}  {K}   \bigg] \\
&=\int dx \bigg[  \frac{p}{D} (Dp)^{q-1} (\frac{K}{p})^2     \bigg]
- \int dx \bigg[  \frac{p}{D} p^{q-1}( D^{q-1} -1 ) (\frac{K}{p})^2  + \frac{u}{D} p^{q-1}{K}   \bigg]\\
&\equiv e_p^{(q)} - h_d^{(q)},
\end{align*}
where the first contribution
\begin{equation} \label{FP e_p}
e_p^{(q)}=
\int dx \bigg[  \frac{p}{D} (Dp)^{q-1} (\frac{K}{p})^2     \bigg]
\geq 0,
\end{equation}
denotes the entropy production rate, which is always non-negative as a manifestation of the
second law of thermodynamics; and the second contribution denotes the heat dissipation rate, or the entropy flux from environment:
\begin{equation}
h_d^{(q)}=
\int dx \bigg[  \frac{p}{D} p^{q-1}( D^{q-1} -1 ) (\frac{K}{p})^2  + \frac{u}{D} p^{q-1}{K}   \bigg].
\end{equation}

{\bf\em House-keeping and non-extensive heats.}
Similarly, the house-keeping heat together with non-extensive heat is identified as the difference between entropy production and free energy dissipation rate, that is,
\begin{align}
e_p^{(q)} - f_d^{(q)}&=
\int dx \frac{p}{D}
\bigg[ (Dp)^{q-1} (\frac{K}{p})^2 -  (\frac{p}{\pi})^{q-1} (\frac{ K }{p} - \frac{K^s}{\pi})^2
     \bigg]
     \equiv Q_{hk}^{(q)} + Q_{ne}^{(q)}, \\
Q_{hk}^{(q)}&= \int dx \bigg[  \frac{p}{D}(D \pi )^{q-1} (\frac{K^s}{\pi})^2  \bigg] \geq 0
, \label{FP Q_hk} \\
Q_{ne}^{(q)}&=\int dx \frac{p}{D}
\bigg[ {(Dp)^{q-1}}(\frac{K}{p})^2 - {(D\pi)^{q-1}}(\frac{K^s}{\pi})^2 -  (\frac{p}{\pi})^{q-1}(\frac{ K }{p} - \frac{K^s}{\pi})^2  \bigg]
 \label{FP ne}
.
\end{align}
Here ${Q}_{hk}^{(q)}$ represents the distance from detailed balance condition, and $Q_{ne}^{(q)}$ arises due to the non-extensibility of Tsallis thermodynamics, which vanishes in the limit of $q\rightarrow 1$,
\begin{equation*}
\lim_{q \rightarrow 1} Q_{ne}^{(q)}=
\int dx  \frac{p}{D}
\bigg[ (\frac{ K }{p})^{2} -  (\frac{K^s}{\pi})^2 -  (\frac{ K }{p} - \frac{K^s}{\pi})^2 \bigg]
= 0 .
\end{equation*}

During above calculations, we use an important relation for F-P equations which is stated through the following lemma.
\begin{lemma} \label{lemma1}
For $q, n \in \mathbb{R}$, such that, $q\neq 0,1$ and $ n \geq 1$, the following integration
\begin{equation}
\int dx  \frac{p}{D}\left(\frac{p}{\pi}\right)^{q-1}
 \left(\frac{ K }{p} - \frac{K^s}{\pi}\right)  \frac{(K^s)^n}{\pi}
=0,
\end{equation}
as long as the probability distribution vanished at the boundary.
\end{lemma}
\begin{proof}
By using $\left(\frac{K}{p} - \frac{K^s}{\pi }\right) = -D \frac{\partial }{\partial x} \ln(\frac{p}{\pi})$, we have
\begin{align*}
&\int dx  \frac{p}{D}(\frac{p}{\pi})^{q-1}
 (\frac{ K }{p} - \frac{K^s}{\pi})  \frac{(K^s)^n}{\pi}
= - \int dx  \frac{\partial }{\partial x} (\frac{p}{\pi})
 (\frac{p}{\pi})^{q-1} {(K^s)^n}
\\
&= - \frac{1}{q} \int dx  \frac{\partial }{\partial x}(\frac{p}{\pi})^{q} {(K^s)^n}
=    \frac{1}{q} \int dx (\frac{p}{\pi})^{q} (K^s)^{n-1}  \frac{\partial }{\partial x} K^s
=0,
\end{align*}
where in the last step we have utilized the divergence-free relation as $\frac{\partial }{\partial x} K^s = 0$ and assumed the probability distribution $p(x,t)$ vanished at the boundary.
\end{proof}

{\bf\em Laws of Tsallis thermodynamics.}
If we further introduce the change rate of excess heat as
\begin{equation}\label{FP excess}
Q_{ex}^{(q)}=\frac{dU^{(q)}}{dt}=
\int dx p^{q-2}K
\bigg[ {\pi}^{2-q} \frac{\partial}{\partial x}(\frac{p}{\pi}) - \frac{\partial p}{\partial x}
     \bigg].
\end{equation}
Then, similar as master equations, laws of Tsallis thermodynamics for F-P equations are readily derived as
\begin{subequations}
\begin{eqnarray}
&&\frac{dS^{(q)}}{dt}+h_{d}^{(q)} =e_p^{(q)} \geq 0, \\
&&\frac{dS^{(q)}}{dt}- Q_{ex}^{(q)} =f_d^{(q)} \geq 0 ,\\
&&Q_{ex}^{(q)}+h_d^{(q)} - Q_{ne}^{(q)}= Q_{hk}^{(q)} \geq 0.
\end{eqnarray}
\end{subequations}
which are also known as three different faces of the second law of thermodynamics according to Ref. \cite{Esposito2010Three}. Again, it should be noted that the non-negativity of $e_p^{(q)} - f_{d}^{(q)}$ is not guaranteed for F-P equations in the framework of Tsallis thermodynamics.

{\bf\em Connections with B-G thermodynamics.}
As ${q \rightarrow 1}$, the entropy production rate, free energy dissipation rate and house-keeping heat of F-P equations become
\begin{align}
& \lim_{q \rightarrow 1} e_p^{(q)}=
\int dx \bigg[  \frac{p}{D} (\frac{K}{p})^2  \bigg]  \geq 0, \label{FP ep rate boltz}\\
&\lim_{q \rightarrow 1} f_d^{(q)} = \int dx \bigg[  \frac{p}{D} (\frac{ K }{p} - \frac{K^s}{\pi})^2
    \bigg] \geq 0, \label{FP fd rate boltz}\\
&\lim_{q \rightarrow 1}Q_{hk}^{(q)}=
\int dx \frac{p}{D}
\bigg[ (\frac{K}{p})^2 - (\frac{ K }{p} - \frac{K^s}{\pi})^2
     \bigg] =
\int dx \frac{p}{D} (\frac{K^s}{\pi})^2 \label{FP Qhk rate boltz} \geq 0,
\end{align}
by using Lemma \ref{lemma1}.
The right-hand side of Eqs. \eqref{FP ep rate boltz}-\eqref{FP Qhk rate boltz} are identified as the corresponding entropy production rate, free energy dissipation rate and house-keeping heat of F-P equations in B-G thermodynamics \cite{van2010,ge2010}.


Furthermore, the limit of heat dissipation rate and change rate of excess heat become, respectively,
\begin{eqnarray}
&&\lim_{q \rightarrow 1} h_d^{(q)}= \int dx \left(K \frac{u}{D}\right), \\
&&\lim_{q \rightarrow 1} Q_{ex}^{(q)}= - \int dx (K \partial_x {\ln \pi}) ,
\end{eqnarray}
which agree with the classical definitions in B-G thermodynamics as expected.

\section{Emergency of a unified formulation in the mathematical limit}
\label{Emergency of a unified formulation in the mathematical limit}

In Ref. \cite{Peng2018The}, a unified B-G thermodynamic formalism has been established for both master equations and F-P equations. Particularly, when a Markov process is restrained to one-step jump, the B-G thermodynamics for the master equations and F-P equations has a one-to-one correspondence. However, in Tsallis thermodynamics, due to the non-extensivity and nonlinearity of Tsallis entropy and free energy, above conclusions may not hold any more. In this part, we are going to show that, within the framework of Tsallis thermodynamics discussed in the last two sections, the free energy dissipation rate $f_d^{(q)}$, entropy production rate $e_p^{(q)}$, and house keeping heat $Q_{hk}^{(q)}$ of master equations with a tridiagonal transition rate matrix can be derived exactly from that of F-P equations; and vice versa. Thus a one-to-one correspondence between the Tsallis thermodynamics of master equations and F-P equations can be rigorously established in mathematics.

\subsection{From F-P equations to master equations by direct discretization}

First, we start with F-P equations with continuous time and state variables, and derive discrete master equations as its approximation.
For notational simplicity, F-P equations in \eqref{FP} is rewritten into a convenient form
\begin{equation}\label{FP1}
\frac{\partial}{\partial t}  p(x,t) =  -\frac{\partial}{\partial x} K(x,t), \quad
K(x,t)={\bar u}(x) p(x,t) - \frac{\partial}{\partial x}\left[ D(x)p(x,t)\right],
\end{equation}
where ${\bar u}(x) \equiv u(x) + {\partial D(x)}/{\partial x}$.
Introduce uniform grids along the x-axis,
\begin{equation}
x=x_i=i \Delta x = i \epsilon, \quad i =0, \pm 1, \pm 2, \cdots.
\end{equation}
By using the forward difference to the drift term $-\frac{\partial }{\partial x}[\bar u(x)p(x,t)]$, and central difference to the diffusion term $\frac{\partial^2}{\partial x^2}[D(x)p(x,t)]$, we discretize the right-hand side of \eqref{FP1} at coordinate $(x_i=i\epsilon, t)$:
\begin{equation*}
\begin{split}
\frac{dp_i(t)}{d t}
=&-\frac{1}{\epsilon} \left[\bar u_{i+1} p_{i+1}(t) - \bar u_{i} p_{i}(t)\right] + \frac{1}{\epsilon^2}\left[D_{i+1} p_{i+1}(t) - 2D_{i} p_{i}(t) + D_{i-1} p_{i-1}(t)\right],
\end{split}
\end{equation*}
where $\bar u_{i} \equiv \bar u(x_i), D_{i} \equiv D(x_i)$, and $p_i(t)$ denotes the probability distribution at $(x_i,t)$. The above equation can be reformulated into the standard form as
\begin{equation}\label{discreteme}
\frac{dp_i(t)}{d t}=
[p_{i-1}(t) w_{i-1, i} - p_{i}(t)w_{i, i-1} ]+
[p_{i+1}(t) w_{i+1, i} - p_{i}(t)w_{i, i+1} ], \quad i =0, \pm 1, \pm 2, \cdots,
\end{equation}
by defining the forward and backward transition rates as
\begin{equation*}
 w_{i-1, i}=\frac{1}{\epsilon^2} D_{i-1},\quad
 w_{i, i-1}=\frac{1}{\epsilon^2}D_{i}-\frac{1}{\epsilon}\bar u_{i},
\end{equation*}
respectively.
Thus we deduce master equations with a tridiagonal transition rate matrix from F-P equations.

Now we are at a position to derive the free energy dissipation rate of master equations from F-P equations. Based on results for Tsallis thermodynamics of F-P equations in Sec. \ref{TsallisFP}, the following theorems can be established. In what follows, for notational simplicity, we will drop the superscript $(q)$ from $f_d^{(q)}$, and use $f_d^M$ and $f_d^F$ to distinguish the free energy dissipation rates of master equations and F-P equations respectively. Similar notations also apply to other quantities without further mention.
\begin{thm}\label{thm3}
In limit of infinitesimal discretization step $\epsilon \rightarrow 0$, the density of free energy dissipation rate \eqref{FP f_d} for F-P equations
\begin{equation}
\lim_{\epsilon \rightarrow 0} \big( \epsilon^{-1} f_d^{F} \big)
=\frac{1}{2}\sum_{i,j}  \phi_{i,j}(t) \Phi_{i,j}(t)
\equiv f_d^M
\geq 0,
\end{equation}
becomes the corresponding result in \eqref{master f_d} for master equations. Here the thermodynamic flux $\phi_{i,j}(t)=p_i w_{i,j} -p_j w_{j,i}$ and force $\Phi_{i,j}(t)=\frac{(p_i/{\pi_i})^{q-1} - (p_j/{\pi_j})^{q-1}}{q - 1}$, with $j=i \pm 1$.
\end{thm}


\begin{thm}\label{thm4}
In the limit of $\epsilon \rightarrow 0$, the densities of entropy production rate \eqref{FP e_p} and house-keeping heat \eqref{FP Q_hk} for F-P equations
\begin{align}
&\lim_{\epsilon \rightarrow 0} \big( \epsilon^{1-2q} e_p^{F} \big)
=\frac{1}{2(q-1)}\sum_{i,j} (p_i w_{i,j} -p_j w_{j,i})  [(p_i w_{i,j})^{q-1} - (p_j w_{j,i})^{q-1}]
\equiv   e_p^M
\geq 0, \\
&\lim_{\epsilon \rightarrow 0} \big( \epsilon^{1-2q} Q_{hk}^{F} \big)
=\frac{1}{2(q-1)}\sum_{i,j}  (p_i w_{i,j} -p_j w_{j,i})
\big[ ({\pi_i w_{ij}})^{q-1} -  ({\pi_j w_{ji}})^{q-1} \big] \\
&~~~~~~~~~~~~~~~~~~~~~~ - \frac{1}{2 q}\sum_{i,j} (\frac{p_i}{\pi_i} - \frac{p_j}{\pi_j}) \big[ ({\pi_i w_{ij}})^{q} -  ({\pi_j w_{ji}})^{q} \big]
\equiv   Q_{hk}^M
\geq 0, \nonumber
\end{align}
lead to, respectively, the corresponding results in Eqs. \eqref{master e_p} and \eqref{master Q_hk} for master equations.
\end{thm}

\subsection{From tridiagonal master equations to F-P equations}
Traditionally, F-P equations can be deduced from master equations by Kramers-Moyal expansion, \textit{i.e.}, by expanding the transition rates and neglecting jump moments of third order and above.
Although Kramers-Moyal expansion offers a straightforward routine from master equations to F-P equations, direct calculations show that thermodynamic elements like $f_d$ are not preserved during this procedure.

To construct a correct limit process, we consider master equations with a tridiagonal transition matrix \cite{van2010}. The elements of $w_{i,j}$ in Eq. \eqref{master_eq} are nonzero only if $j=i \pm 1$ or $j=i$ with $w_{i,i} = - w_{i,i-1}-w_{i, i+1}$.
Then master equation \eqref{master_eq} is written as follows
\begin{equation}\label{master2}
\frac{dp_i}{dt}= (p_{i+1} w_{i+1,i} - p_{i} w_{i,i-1}) - (p_{i} w_{i,i+1} - p_{i-1} w_{i-1,i}).
\end{equation}
To make the state variables to be continuous, let us introduce
\begin{align} \label{rescale}
&x=i\epsilon,\quad  p(x,t)=p_i(t), \quad
2D(x)=(w_{i-1,i}+w_{i,i-1})\epsilon^2, \quad u(x)=(w_{i-1,i} - w_{i,i-1})\epsilon,
\end{align}
with a small parameter $\epsilon$ ($|\epsilon|\ll1$)  to measure the relative importance of fluctuations, as ${1}/{\epsilon}$ represents the system volume or number of particles.
Taking the limit $\epsilon \rightarrow 0$, Eq. \eqref{master2} becomes
\begin{align}\label{FP2}
\frac{\partial p}{\partial t}
&= \lim_{\epsilon \rightarrow 0}
 \partial_x \bigg\{\epsilon p(\frac{1}{\epsilon^2}D -\frac{1}{2\epsilon} u) - \epsilon[p- \epsilon \partial_x p +\mathcal{O}({\epsilon^2})] (\frac{1}{\epsilon^2}D + \frac{1}{2\epsilon}u )
\bigg\}  +\mathcal{O}({\epsilon})
\nonumber \\
&= - \partial_x (up -D\partial_x p)
\equiv - \partial_x K(x,t),
\end{align}
which is the classical F-P equation with the probability flux $K=up -D\partial_x p$.

We have shown that F-P equations can be derived rigorously from the tridiagonal master equations.
Starting from this point, we can further prove that the free energy dissipation rate $f_d^{M}$ for master equations converges automatically to $f_d^{F}$ for the corresponding F-P equations when $\epsilon  \rightarrow 0$.

\begin{thm}\label{thm2}
In the limit $\epsilon \rightarrow 0$, the density of free energy dissipation rate \eqref{master f_d} for master equations
\begin{equation}
\lim_{\epsilon \rightarrow 0} \big( \epsilon f_d^{M} \big)
= - \int dx  \frac{K}{q-1}  \frac{\partial}{\partial x} \big[ (\frac{p}{\pi})^{q-1}\big]
\equiv  f_d^{F} \geq 0,
\end{equation}
becomes the free energy dissipation rate \eqref{FP f_d} for F-P equations.
\end{thm}

\begin{thm}\label{thm4}
In the limit $\epsilon \rightarrow 0$, the densities of entropy production rate \eqref{master e_p} and house-keeping heat \eqref{master Q_hk} for master equations
\begin{align}
&\lim_{\epsilon \rightarrow 0}  \big( \epsilon^{2q-1}e_p^M  \big)
=\int dx \bigg[  \frac{p}{D} (Dp)^{q-1} (\frac{K}{p})^2     \bigg]
\equiv e_p^F
\geq 0, \\
&\lim_{\epsilon \rightarrow 0}  \big( \epsilon^{2q-1} Q_{hk}^M  \big)
=\int dx \bigg[  \frac{p}{D}
(D\pi)^{q-1} (\frac{K^s}{\pi})^{2}
\bigg]
\equiv Q_{hk}^F
\geq 0,
\end{align}
become, respectively, the entropy production rate \eqref{FP e_p} and house-keeping heat \eqref{FP Q_hk} for F-P equations.
\end{thm}

\begin{rem}
Generally speaking, there are many different coarse-graining ways from a description of lower level to that of higher level,
for instance the Kramers-Moyal expansion and canonical form expansion of master equations in our case.
We notice that the canonical form expansion \cite{van1983, Peng2018The} well preserves the free energy dissipation rate during the coarse-graining procedure, while the Kramers-Moyal expansion fails. Together with the results in Ref. \cite{Peng2018The}, we may conclude that canonical form expansion is more appropriate than Kramers-Moyal expansion from the viewpoint of nonequilibrium thermodynamics.
\end{rem}

\section{Applications}
\subsection{Two-state chemical reactions}
We firstly consider $N$ particles occupying two giving states $1$ and $2$ \cite{qian2007phosphorylation}, like protein folding and unfolding, DNA phosphorylation and dephosphorylation, gene on and off. With constant jumping rates  $\mu$ and $\lambda$, the probability of the system with $i$ particles at state $1$ and $(N-i)$ particles at state $2$ evolves according to the following master equations,
\begin{equation}
\frac{dp(i,N-i)}{dt} = \sum_{i=0}^N \left[\mu p(i+1,N-i-1)-(\mu+\lambda)p(i,N-i)+\lambda p(i-1,N-i+1)\right],
\end{equation}
where $i=0,1,\cdots,N$. 

\begin{figure}[!htp]
\centering
\includegraphics[width=0.8\linewidth]{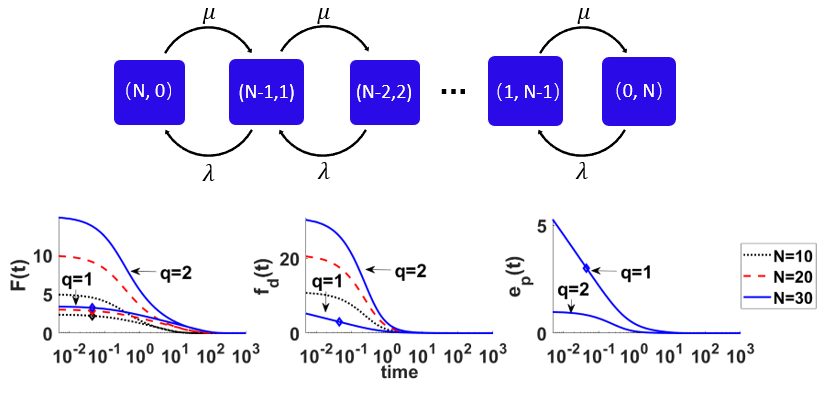}
\caption{Comparison of Tsallis thermodynamics ($q=2$) and B-G thermodynamics ($q=1$) on their system size dependence. The initial state is prepared at $p(N,0)=1$, which finally reaches a uniform distribution.}
\label{fig.master}
\end{figure}

Here we particularly focus on the dependence of Tsallis thermodynamics and B-G thermodynamics with respect to the system size. As illustrated through Fig. \ref{fig.master}, Tsallis thermodynamics with $q>1$ generally shows a more apparent dependence on particle numbers $N$ than the B-G thermodynamics. For example, in B-G thermodynamics both $f_d$ and $e_p$ scale independent of the system size, but in Tsallis thermodynamics $f_d$ shows a size dependence while $e_p$ does not. This fact highlights the distinction of Tsallis thermodynamics on its non-extensive nature from B-G thermodynamics.

\subsection{Optical lattices}
In the second example, we look into optical lattices described by a Fokker-Planck equation \eqref{FP} for the Wigner function, where
\begin{equation}\label{op.la}
u(x)=-\frac{\alpha x}{1+(x/x_c)^2}, \quad D(x)=D_0+\frac{D_1}{1+(x/x_c)^2}.
\end{equation}
Note the Wigner function is actually in the momentum space, though we keep using variable $x$ in accordance with previous notations in this paper. Lutz firstly pointed out the above F-P equation admits Tsallis statistics as a stationary solution \cite{lutz2003anomalous},
\begin{equation}
\pi(x)=C\bigg[1+\frac{D_0}{D_0+D_1}\left(\frac{x}{x_c}\right)^2\bigg]^{1/(1-q)},
\end{equation}
where $q=1+2D_0/(\alpha x_c^2)$. So the optical lattice effectively maximizes the Tsallis entropy in its steady state.

\begin{figure}[!htp]
\centering
\includegraphics[width=\linewidth]{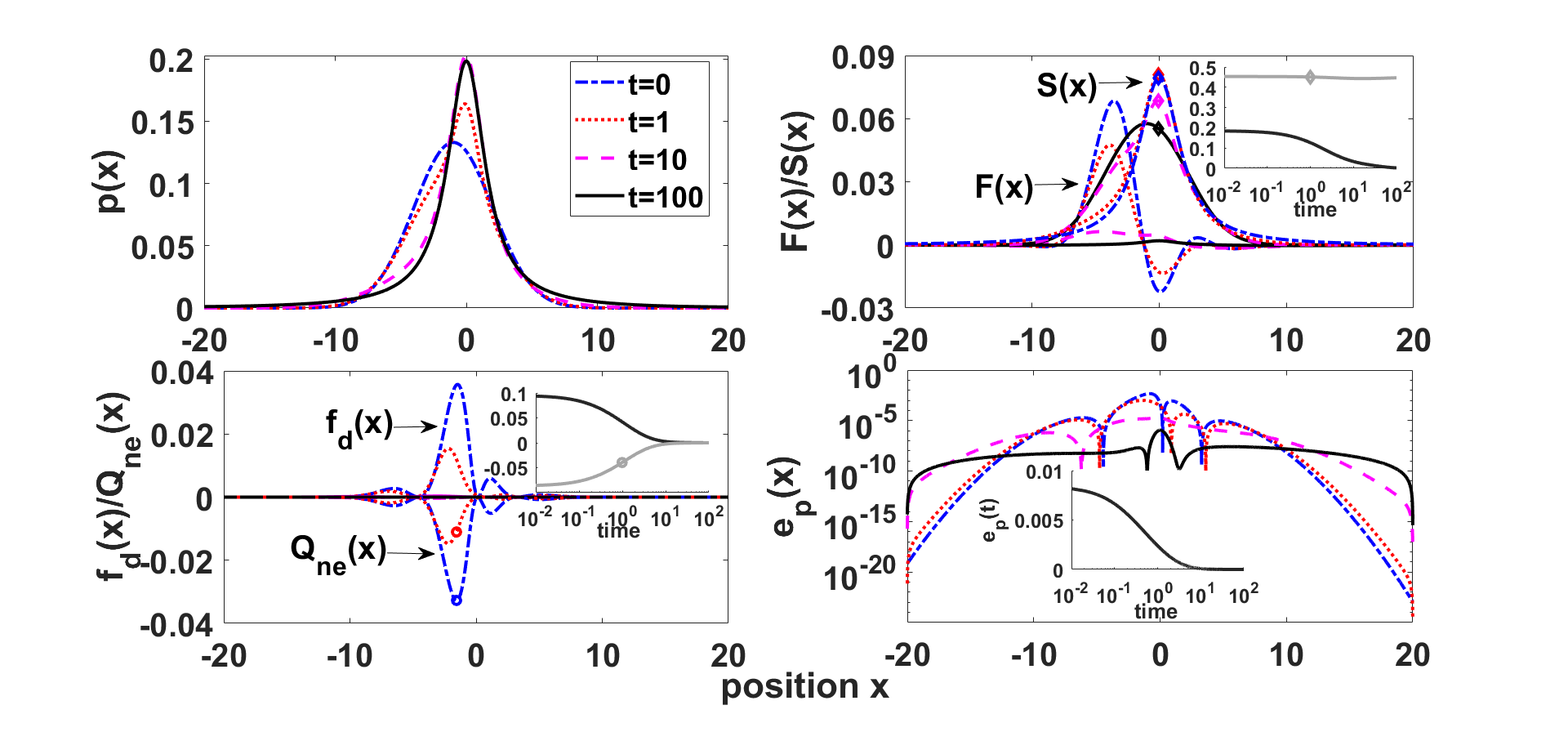}
\caption{The solution and Tsallis thermodynamics for optical lattices modeling by F-P equations with $\alpha=D_1=x_c=1$ and $D_0=1/2$. Spatial distributions of each quantity (before integration) are illustrated at four representative time points, while their overall time evolutions (after integration) are drawn in the insets correspondingly. }
\label{fig.FP}
\end{figure}

Typical temporal and spatial behaviors of the Tsallis thermodynamics for optical lattices with $q=2$ are explored in Fig. \ref{fig.FP}. Starting from a Gaussian distribution with center a little shifted from zero, solutions of F-P equations in \eqref{FP} and \eqref{op.la} gradually approach to the stationary Tsallis statistics as time evolves. During this procedure, the free energy $F^{(q)}$, free energy dissipation rate $f_d^{(q)}$ and entropy production rate $e_p^{(q)}$ all converge to zero from the positive side as our results have pointed out. A spatial plot on these quantities before integration further shows that their non-negativeness also holds from point to point, expect for the free energy. The non-extensive heat $Q_{ne}^{(q)}$ in this case is almost as large as the free energy dissipation rate but with a minus sign, which indicates the importance of this quantity in maintaining a correct form of Tsallis thermodynamics.

\section{Conclusion}
\label{Conclusion}

It is well-known that the B-G thermodynamics is inadequate for small systems, in which the non-extensivity plays a fundamental role. Parallel to the B-G statistical thermodynamics, this paper aims to establish a rigorous and self-consistent thermodynamic formalism of master and Fokker-Planck equations with non-extensive Tsallis entropy and free energy. Interestingly, the conventional thermodynamics with B-G entropy turns to be a special case of our study in the limit of $q\rightarrow 1$.

To be concrete, we first derived the Tsallis thermodynamics for Markov processes described by master equations, including the entropy production rate $e_p^{(q)}$, free energy dissipation rate $f_d^{(q)}$, house-keeping heat $Q_{hk}^{(q)}$, heat dissipation rate $h_d^{(q)}$, excess heat exchange rate $Q_{ex}^{(q)}$ and the non-extensive heat $Q_{ne}^{(q)}$, as well as three balance equations involving above quantities as a manifestation of ``three faces of the second law of thermodynamics'' according to Esposito \textit{et al.}\cite{esposito2010}. Then, we follow the same routine and build the Tsallis thermodynamics for F-P equations.
Finally, to justify above formulations, we deduced the Tsallis thermodynamics of master equations from the discretized F-P equations when the discretization step size goes to zero. A similar one-to-one correspondence emerges from master equations to F-P equations too, when the Markov process is restrained to one-step jump.

Compared to results for B-G thermodynamics, a direct consequence of the non-extensivity of Tsallis free energy is the emergency of a non-local Onsager's force (see Eq. (\ref{O_force})).
Besides, the difference between the entropy production rate and free energy dissipation rate is not positive any more. Actually, as we have shown, only a part of it gives the correct house-keeping heat, which is always non-negative and vanishes if and only if at the steady state. The reminding part characterizes the non-extensivity of the system and is thus named as the non-extensive heat $Q_{ne}^{(q)}$, since $\lim_{q\rightarrow 1} Q_{ne}=0$.
Furthermore, we have also shown the power of coarse-graining procedure or correct mathematical limit process in discovering new thermodynamic expressions and relations, exploring underlying novel connections and correspondence between different levels of descriptions of the same system, which turns out to both mathematically critical and physically insightful in the study of non-equilibrium thermodynamics.

\section*{acknowledgment}
L.H. acknowledged the financial supports from the National Natural Science Foundation of China (Grants 21877070) and Tsinghua University Initiative Scientific Research Program (Grants 20151080424).

\section*{Appendix: Rigorous proofs on Tsallis thermodynamics of master equations and F-P equations}
(1) {\bf\em Proof of Theorem 4.1:}
The free energy dissipation rate for F-P equations is given by
\begin{align*}
f_d^F
= - \int {dx} \big[ K(x,t)\frac{1}{q-1} \frac{\partial}{\partial x}(\frac{p}{\pi})^{q-1} \big].
\end{align*}
By Taylor series expansion, we discretize the integral function with respect to $x=i \epsilon$ to obtain
\begin{align*}
f_d^F(\epsilon)
&=- \sum_{i}  \frac{\epsilon}{q-1}
\bigg[\bar u_{i}p_{i} - \frac{1}{\epsilon}({D_{i+1}p_{i+1} - D_{i}p_{i}}) + \mathcal{O}(\epsilon) \bigg]
\bigg\{
\frac{1}{\epsilon }\big[  (\frac{p_{i+1}}{\pi_{i+1}})^{q-1}  - (\frac{p_{i}}{\pi_{i}})^{q-1} \big]  + \mathcal{O}(\epsilon) \bigg\}\\
&=\epsilon\sum_{i}
\bigg[ \frac{D_{i+1}p_{i+1} - D_{i}p_{i} -\epsilon \bar u_{i+1}p_{i+1} }{\epsilon^2} +\mathcal{O}( 1 ) \bigg]
 \bigg\{  \frac{1}{q-1}\big[(\frac{p_{i+1}}{\pi_{i+1}})^{q-1}  - (\frac{p_{i}}{\pi_{i}})^{q-1}\big] +\mathcal{O}(\epsilon^2) \bigg\}   \\
&\equiv \epsilon \sum_{i}  I_1(i,t)I_{2}(i,t)  ,
\end{align*}
where the summand is denoted as $I_1(i,t)I_2(i,t)$. The first part and second part read separately as
\begin{align*}
I_1(i,t)
&\equiv
\frac{1}{\epsilon^2}  \big[ D_{i+1}p_{i+1} - D_{i}p_{i} -\epsilon \bar u_{i+1}p_{i+1} \big]+\mathcal{O}(1)
 \\&
=- (p_i w_{i,i+1}-p_{i+1} w_{i+1,i} ) + \mathcal{O}(1),\\
I_{2}(i,t)
&\equiv
\frac{1}{ (q-1)} \bigg[ -(\frac{p_{i}}{\pi_{i}})^{q-1} + (\frac{p_{i+1}}{\pi_{i+1}})^{q-1}\bigg] +\mathcal{O}(\epsilon^2) .
\end{align*}
We notice that the leading terms of $I_1(i,t)$ and $I_{2}(i,t)$ are of orders $\mathcal{O}(1/\epsilon)$ and $\mathcal{O}(\epsilon)$, respectively.
Direct calculations indicate that,
\begin{align*}
&\lim_{\epsilon \rightarrow 0} \big( \epsilon^{-1} f_d^{F} \big)\\
=& \lim_{\epsilon\rightarrow 0}
   \sum_{i}  \big[ (p_i w_{i,i+1}-p_{i+1} w_{i+1,i})+\mathcal{O}(1) \big]
              \cdot \bigg\{
 \frac{1}{ (q-1) } \big[ (\frac{p_{i}}{\pi_{i}})^{q-1} - (\frac{p_{i+1}}{\pi_{i+1}})^{q-1} \big] +\mathcal{O}(\epsilon^2)
 \bigg\} \\
=&  \sum_{i}  \frac{1}{q-1} (p_i w_{i,i+1}-p_{i+1} w_{i+1,i})  \big[(\frac{p_{i}}{\pi_{i}})^{q-1} - (\frac{p_{i+1}}{\pi_{i+1}})^{q-1} \big] \\
=& \frac{1}{2(q-1)}\sum_{i,j=i\pm 1}  (p_i w_{i,j}-p_{j} w_{j,i})  \big[(\frac{p_{i}}{\pi_{i}})^{q-1} - (\frac{p_{j}}{\pi_{j}})^{q-1} \big],
\end{align*}
where in the last step we have translated the summation index from $i$ to $(i-1)$.
This completes the proof.\bigskip

(2) {\bf\em Proof of Theorem 4.2:}
Discretizing the entropy production rate $e_p^F$ for F-P equations, one has
\begin{align*}
&e_p^F(\epsilon)
=\epsilon \sum_{i} (D_i p_i)^{q-2}
\bigg[\bar u_{i}p_{i} - \frac{1}{\epsilon}({D_{i+1}p_{i+1} - D_{i}p_{i}}) + \mathcal{O}(\epsilon) \bigg]^2
\\
=&\frac{1}{\epsilon}\sum_i \big[ D_{i+1}p_{i+1} - D_{i}p_{i} -\epsilon \bar u_{i+1}p_{i+1} +\mathcal{O}(\epsilon^2) \big]\big[ \frac{ D_{i+1}p_{i+1} - D_{i}p_{i} -\epsilon \bar u_{i+1}p_{i+1} }{D_i p_i} +\mathcal{O}(\epsilon^2) \big] (D_i p_i)^{q-1} \\
\equiv& \epsilon \sum_{i}  I_1(i,t)I_{3}(i,t).
\end{align*}
Here
\begin{align*}
I_1(i,t)
&\equiv
\frac{1}{\epsilon^2}  \big[ D_{i+1}p_{i+1} - D_{i}p_{i} -\epsilon \bar u_{i+1}p_{i+1} +\mathcal{O}(\epsilon^2) \big]
=- (p_i w_{i,i+1}-p_{i+1} w_{i+1,i} ) + \mathcal{O}(1),\\
I_{3}(i,t)
&\equiv
\big[ \frac{ D_{i+1}p_{i+1} - D_{i}p_{i} -\epsilon \bar u_{i+1}p_{i+1} }{D_i p_i} +\mathcal{O}(\epsilon^2) \big] (D_i p_i)^{q-1} \\
&=\frac{1}{q-1}\big[ (\frac{ D_{i+1}p_{i+1}  -  \epsilon \bar u_{i+1}p_{i+1} }{D_i p_i})^{q-1} -1 +\mathcal{O}(\epsilon^2) \big] (D_i p_i)^{q-1} \\
&= \frac{\epsilon^{2(q-1)}}{q-1}
\big[(p_{i+1} w_{i+1,i})^{q-1} - (p_i w_{i,i+1})^{q-1} +\mathcal{O}(\epsilon^2) \big]
,
\end{align*}
where we have used the relation $\frac{1}{q-1} [(1+\zeta)^{q-1} -1] =\zeta +\mathcal{O}({\zeta^2}) $ for $|\zeta|\ll1$ in the second step during the derivation of $I_{3}$.
Therefore, the limit of entropy production rate becomes
\begin{align*}
\lim_{\epsilon \rightarrow 0} \big( \epsilon^{1-2q} e_p^{F} \big)
&=\sum_{i}
 \frac{1}{q-1}(p_i w_{i,i+1}-p_{i+1} w_{i+1,i})
\big[  (p_i w_{i,i+1})^{q-1} -  (p_{i+1} w_{i+1,i})^{q-1}  \big]\\
&= \frac{1}{2(q-1)}\sum_{i,j}
(p_i w_{i,j}-p_{i+1} w_{j,i})
\big[  (p_i w_{i,j})^{q-1} -  (p_{j} w_{j,i})^{q-1}  \big] ,
\end{align*}
which is exactly the expression of $e_p^M$ in Eq. \eqref{master e_p} for master equations with $j= i\pm 1$.

As to the house-keeping heat, we can rewrite $Q_{hk}^F$ in Eq. \eqref{FP Q_hk}  into an alternative form:
\begin{align*}
Q_{hk}^F
&= \int dx \bigg[  \frac{p}{D}(D \pi )^{q-1} (\frac{K^s}{\pi})^2  \bigg]
=\int dx \bigg\{ [K -(K-\frac{p}{\pi}K^s)] K^s (D \pi )^{q-2}  \bigg\}\\
&=\int dx \bigg\{ [K + D\pi \frac{\partial}{\partial x}(\frac{p}{\pi}) ] K^s (D \pi )^{q-2}  \bigg\}
.
\end{align*}
Discretizing $Q_{hk}^F$ by $x=i \epsilon$, one has
\begin{align*}
&Q_{hk}^F(\epsilon)
=\epsilon \sum_{i}
\bigg[\bar u_{i}p_{i} - \frac{1}{\epsilon}({D_{i+1}p_{i+1} - D_{i}p_{i}})
+ \frac{1}{\epsilon} D_i \pi_i (\frac{p_{i+1}}{\pi_{i+1}} - \frac{p_{i}}{\pi_{i}})
+ \mathcal{O}(\epsilon) \bigg] \\
& ~~~~~~~ \cdot \bigg[ \bar u_{i}\pi_{i} - \frac{1}{\epsilon}({D_{i+1}\pi_{i+1} - D_{i}\pi_{i}})
+ \mathcal{O}(\epsilon) \bigg] (D_i \pi_i)^{q-2}
\\
=& \epsilon \sum_{i}
 \bigg[ \frac{D_{i+1}p_{i+1} - D_{i}p_{i} -\epsilon \bar u_{i+1}p_{i+1}}{\epsilon^2 }  + \mathcal{O}(1) \bigg]
    \big[ D_{i+1}\pi_{i+1} - D_{i}\pi_{i} -\epsilon \bar u_{i+1}\pi_{i+1} + \mathcal{O}(\epsilon^2) \big]  (D_i \pi_i)^{q-2} \\
& - \epsilon \sum_{i}
\bigg[  \frac{D_i \pi_i }{\epsilon^2 } (\frac{p_{i+1}}{\pi_{i+1}} - \frac{p_{i}}{\pi_{i}}) + \mathcal{O}(1)  \bigg]
   \big[ D_{i+1}\pi_{i+1} - D_{i}\pi_{i} -\epsilon \bar u_{i+1}\pi_{i+1} + \mathcal{O}(\epsilon^2) \big]  (D_i \pi_i)^{q-2}
 \\
\equiv  & \epsilon \sum_{i}  [ I_1(i,t)I_{4}(i,t) -  I_{5}(i,t) ]
.
\end{align*}
Similarly,
\begin{align*}
 I_{4}
&\equiv
 \big[ \frac{D_{i+1}\pi_{i+1} - D_{i}\pi_{i} -\epsilon \bar u_{i+1}\pi_{i+1}}{D_i \pi_i} +\mathcal{O}(\epsilon^2) \big] (D_i \pi_i)^{q-1} \\
& =  \frac{\epsilon^{2(q-1)}}{q-1}
\big[ (\pi_{i+1} w_{i+1,i})^{q-1}  -  (\pi_i w_{i,i+1})^{q-1}  +  \mathcal{O}(\epsilon^2) \big] ,
\\
I_{5}
&\equiv
\frac{1}{\epsilon^2 }
\big[ (\frac{p_{i+1}}{\pi_{i+1}} - \frac{p_{i}}{\pi_{i}}) + \mathcal{O}(\epsilon^2)  \big]
 \big[ \frac{D_{i+1}\pi_{i+1} - D_{i}\pi_{i} -\epsilon \bar u_{i+1}\pi_{i+1}}{D_i \pi_i}  +\mathcal{O}(\epsilon^2) \big]  (D_i \pi_i)^{q}  \\
&=\frac{\epsilon^{2(q-1)}}{q}
\big[ (\frac{p_{i+1}}{\pi_{i+1}} - \frac{p_{i}}{\pi_{i}}) + \mathcal{O}(\epsilon^2)  \big]
\big[  (\pi_{i+1} w_{i+1,i})^{q} -  (\pi_i w_{i,i+1})^{q}  +\mathcal{O}(\epsilon^2) \big]
.
\end{align*}
Finally, we arrive at the house-keeping heat of master equations with a tridiagonal transition rate matrix as
\begin{align*}
&\lim_{\epsilon \rightarrow 0} \big( \epsilon^{1-2q}Q_{hk}^{F} \big)
=\frac{1}{2(q-1)}\sum_{i,j}  (p_i w_{i,j} -p_j w_{j,i})
\big[ ({\pi_i w_{i,j}})^{q-1} -  ({\pi_j w_{j,i}})^{q-1} \big] \\
&~~~~~~~~~~~~~~~~~~~~~~ - \frac{1}{ 2 q }\sum_{i,j} (\frac{p_i}{\pi_i} - \frac{p_j}{\pi_j}) \big[ ({\pi_i w_{i,j}})^{q} -  ({\pi_j w_{j,i}})^{q} \big],
\end{align*}
based on the symmetry of summands with respect to the indices $i$ and $j=i \pm 1$.\bigskip

(3) {\bf\em Proof of Theorem 4.3:}
Recalling the free energy dissipation rate for master equations
\begin{equation*}
f_d^M(t)
=\frac{1}{2} \sum_{i,j} \phi_{i,j}(t) \Phi_{i,j}(t) \geq 0,
\end{equation*}
with $j=i\pm 1$. We take $j=i-1$ as an example. The case $j=i+1$ is the same, since $\sum_{i} \phi_{i,i+1} \Phi_{i,i+1}= \sum_{i} \phi_{i-1,i} \Phi_{i-1,i} = \sum_{i} \phi_{i,i-1} \Phi_{i,i-1}$ due to the symmetry of indices and reflecting boundary conditions.

First, we substitute the relations \eqref{rescale} into the flux
\begin{align*}
\phi_{i,i-1}
=& p_{i} w_{i,i-1} - p_{i-1} w_{i-1,i} \\
=& p(\frac{1}{\epsilon^2}D -\frac{1}{2\epsilon} u) -  [p- \epsilon \partial_x p +\mathcal{O}({\epsilon^2})] (\frac{1}{\epsilon^2}D + \frac{1}{2\epsilon}u ) \\ 
=&- \frac{1}{\epsilon}(up -D\partial_x p) + \mathcal{O}(1)
\equiv - \frac{1}{\epsilon} K  + \mathcal{O}(1).
\end{align*}
Second, the force term is treated as
\begin{align*}
\Phi_{i,i-1}
=& \frac{1}{q-1} \big[ (\frac{p_{i}}{\pi_{i}})^{q-1} -  (\frac{p_{i-1}}{\pi_{i-1}})^{q-1} \big] =\frac{\epsilon}{q-1} \big[ \frac{\partial}{\partial x}  (\frac{p}{\pi})^{q-1}  +\mathcal{O}({\epsilon}) \big]
.
\end{align*}
Recalling the relation $\sum_i = \frac{1}{\epsilon} \int dx $, one has the density of free energy dissipation rate for master equations
\begin{equation*}
\lim_{\epsilon \rightarrow 0} \big( \epsilon f_d^{M} \big)
= - \int dx  \frac{K}{q-1}  \frac{\partial}{\partial x} \big[ (\frac{p}{\pi})^{q-1}\big]
\equiv  f_d^{F} \geq 0.
\end{equation*}\bigskip

(4) {\bf\em Proof of Theorem 4.4:}
Denote the entropy production rate for master equations as
\begin{align*}
& e_p^M
=\frac{1}{2(q-1)}\sum_{i,j} (p_i w_{i,j} -p_j w_{j,i})  [(p_i w_{i,j})^{q-1} - (p_j w_{j,i})^{q-1}]
\equiv \frac{1}{2 }\sum_{i,j} \phi_{i,j} \Psi_{i,j},
\end{align*}
with
\begin{align*}
&\Psi_{i,j}
=\frac{1}{q-1} [(p_i w_{i,j})^{q-1} - (p_j w_{j,i})^{q-1}].
\end{align*}
Choosing $j=i-1$ and substituting the relations \eqref{rescale} into the above equation, one deduces that
\begin{align*}
&\Psi_{i,i-1}
=\frac{1}{q-1} [(p_i w_{i,i-1})^{q-1} - (p_{i-1} w_{i-1,i})^{q-1}] \\
=& \frac{1}{q-1}  \bigg\{ \big[ p(\frac{1}{\epsilon^2}D -\frac{1}{2\epsilon} u)  \big]^{q-1} -  \big[ p- \epsilon \partial_x p +\mathcal{O}({\epsilon^2})\big]^{q-1} (\frac{1}{\epsilon^2}D + \frac{1}{2\epsilon}u )^{q-1}
\bigg\}
\\ 
=& \frac{1}{\epsilon^{2(q-1)}}  \frac{(Dp)^{q-1}}{q-1}  \bigg\{ \big( 1 - \epsilon \frac{u}{2D}  \big)^{q-1} -  \big[ 1 + \epsilon \frac{u}{2D}  - \epsilon  \partial_x (\ln p ) +\mathcal{O}({\epsilon^2})\big]^{q-1}
\bigg\}
\\
=&-\frac{1}{\epsilon^{2q-3}} (Dp)^{q-2} [ K +\mathcal{O}({\epsilon})]
.
\end{align*}
Combining the flux $\phi_{i,i-1}$ and the associated force $\Psi_{i,i-1}$, we have
\begin{align}
&\lim_{\epsilon \rightarrow 0}  \big( \epsilon^{2q-1}e_p^M  \big)
=\int dx (Dp)^{q-2} K^2
\geq 0,
\end{align}
which is exactly the definition of $e_p^F$ in \eqref{FP e_p} for F-P equations.

The house-keeping heat could be treat analogously. Based on Eq. \eqref{master Q_hk}, $Q_{hk}^M $ is rewritten as
\begin{align*}
Q_{hk}^M=
&\sum_{i} \frac{1}{q-1} (p_i w_{i,i-1} - p_{i-1} w_{i-1,i})[(\pi_i w_{i,i-1})^{q-1} - (\pi_{i-1} w_{i-1,i})^{q-1}] \\
&-\sum_{i} \frac{1}{ q} (\frac{p_i}{\pi_i} - \frac{p_{i-1}}{\pi_{i-1}}) \big[ ({\pi_i w_{i,{i-1}}})^{q} -  ({\pi_{i-1} w_{{i-1},i}})^{q} \big]\\
\equiv
& \sum_{i} \phi_{i,i-1} \bar\Psi_{i,i-1} -  \sum_{i} \Delta_{i,i-1}.
\end{align*}
We notice that the first term of $Q_{hk}^M $ is the product of the flux $\phi_{i,i-1}$ and corresponding force
\begin{align*}
&\bar\Psi_{i,i-1}
\equiv \frac{1}{q-1}
\big[ ({\pi_i w_{i,i-1}})^{q-1} -  ({\pi_{i-1} w_{{i-1}, i}})^{q-1} \big]
=-\frac{1}{\epsilon^{2q-3}} (D\pi)^{q-2} [ K^s +\mathcal{O}({\epsilon})],
\end{align*}
while the second terms is
\begin{align*}
\Delta_{i,i-1}
\equiv &
\frac{1}{ q} (\frac{p_i}{\pi_i} - \frac{p_{i-1}}{\pi_{i-1}}) \big[ ({\pi_i w_{i,{i-1}}})^{q} -  ({\pi_{i-1} w_{{i-1},i}})^{q} \big]\\
=& \frac{1}{q} \big[  \epsilon \frac{\partial}{\partial x}  (\frac{p}{\pi})  +\mathcal{O}({\epsilon}^2)  \big]
\bigg\{ \big[ \pi(\frac{1}{\epsilon^2}D -\frac{1}{2\epsilon} u)  \big]^{q} -  \big[ \pi- \epsilon \partial_x \pi +\mathcal{O}({\epsilon^2})\big]^{q} (\frac{1}{\epsilon^2}D + \frac{1}{2\epsilon}u )^{q}
\bigg\}
\\
=&-\frac{1}{\epsilon^{2q-2}} (D\pi)^{q-1}  \big[\frac{\partial}{\partial x}  (\frac{p}{\pi})  +\mathcal{O}({\epsilon})\big]
[ K^s +\mathcal{O}({\epsilon})]
.
\end{align*}
As a result, in the limit of $\epsilon \rightarrow 0$, the density of house-keeping heat of master equations becomes
\begin{align*}
& \lim_{\epsilon \rightarrow 0}  \big( \epsilon^{2q-1} Q_{hk}^M  \big)
=\int dx \bigg[
(D\pi)^{q-2} K  + (D\pi)^{q-1} \frac{\partial}{\partial x}(\frac{p}{\pi})
\bigg]  K^s \\
&=\int dx \bigg[  \frac{p}{D}
(D\pi)^{q-1} (\frac{K^s}{\pi})^{2}
\bigg]
\geq 0,
\end{align*}
which emerges as $Q_{hk}^F$ for F-P equations.
This completes our proof.

\bibliographystyle{unsrt}
\bibliography{fp}

\begin{thebibliography}{10}

\bibitem{ge2010}
Hao Ge and Hong Qian.
\newblock Physical origins of entropy production, free energy dissipation, and
  their mathematical representations.
\newblock {\em Physical Review E}, 81(5):051133, 2010.

\bibitem{esposito2010}
Massimiliano Esposito and Christian Van~den Broeck.
\newblock Three faces of the second law. i. master equation formulation.
\newblock {\em Physical Review E}, 82(1):011143, 2010.

\bibitem{Ge2009Extended}
Hao Ge.
\newblock Extended forms of the second law for general time-dependent
  stochastic processes.
\newblock {\em Physical Review E}, 80(2):021137, 2009.

\bibitem{van2010}
Christian Van~den Broeck and Massimiliano Esposito.
\newblock Three faces of the second law. ii. {Fokker-Planck} formulation.
\newblock {\em Physical Review E}, 82(1):011144, 2010.

\bibitem{Hill1963Thermodynamics}
Terrell~L. Hill.
\newblock {\em Thermodynamics of small systems}.
\newblock Dover, New York, 1963.

\bibitem{Abe2001Nonextensive}
Sumiyoshi Abe and Yuko Okamoto.
\newblock {\em Nonextensive statistical mechanics and its applications}, volume
  560.
\newblock Springer Science \& Business Media, 2001.

\bibitem{tsallis1988}
Constantino Tsallis.
\newblock Possible generalization of {Boltzmann-Gibbs} statistics.
\newblock {\em Journal of Statistical Physics}, 52(1):479--487, 1988.

\bibitem{Renyi1959}
A.~R{\'e}nyi.
\newblock On the dimension and entropy of probability distributions.
\newblock {\em Acta Mathematica Academiae Scientiarum Hungarica},
  10(1):193--215, Mar 1959.

\bibitem{Renyi1970Probability}
A.~R{\'e}nyi.
\newblock {\em Probability Theory}.
\newblock North Holland, Amsterdam, 1970.

\bibitem{Kaniadakis2001Nonlinear}
G.~Kaniadakis.
\newblock Non-linear kinetics underlying generalized statistics.
\newblock {\em Physica A: Statistical Mechanics and its Applications},
  296(3-4):405--425, 2001.

\bibitem{Boon2005Special}
Jean~Pierre Boon and Constantino Tsallis.
\newblock Special issue overview nonextensive statistical mechanics: new
  trends, new perspectives.
\newblock {\em Europhysics News}, 36(6):185--186, 2005.

\bibitem{beck2009generalised}
Christian Beck.
\newblock Generalised information and entropy measures in physics.
\newblock {\em Contemporary Physics}, 50(4):495--510, 2009.

\bibitem{daniels2004defect}
Karen~E Daniels, Christian Beck, and Eberhard Bodenschatz.
\newblock Defect turbulence and generalized statistical mechanics.
\newblock {\em Physica D: Nonlinear Phenomena}, 193(1-4):208--217, 2004.

\bibitem{lutz2003anomalous}
Eric Lutz.
\newblock Anomalous diffusion and tsallis statistics in an optical lattice.
\newblock {\em Physical Review A}, 67(5):051402, 2003.

\bibitem{borland2002option}
Lisa Borland.
\newblock Option pricing formulas based on a non-gaussian stock price model.
\newblock {\em Physical Review Letters}, 89(9):098701, 2002.

\bibitem{amari2007methods}
Shun-ichi Amari and Hiroshi Nagaoka.
\newblock {\em Methods of information geometry}, volume 191.
\newblock American Mathematical Soc., 2007.

\bibitem{kac2001quantum}
Victor Kac and Pokman Cheung.
\newblock {\em Quantum calculus}.
\newblock Springer Science \& Business Media, 2001.

\bibitem{hobson1969}
A.~Hobson.
\newblock A new theorem of information theory.
\newblock {\em Journal of Statistical Physics}, 1(3):383--391, 1969.

\bibitem{Shore1980Axiomatic}
John~E Shore and Rodney~W Johnson.
\newblock Axiomatic derivation of the principle of maximum entropy and the
  principle of minimum cross-entropy.
\newblock {\em Information Theory IEEE Transactions on}, 26(1):26--37, 1980.

\bibitem{shiino1998}
Masatoshi Shiino.
\newblock H-theorem with generalized relative entropies and the {Tsallis}
  statistics.
\newblock {\em Journal of the Physical Society of Japan}, 67(11):3658--3660,
  1998.

\bibitem{Khinchin1953The}
A.I. Khinchin.
\newblock The entropy concept in probability theory. {English} translation in
  {Khinchin, A.I.} (1957) { {Mathematical Foundations of Information Theory}}.
  {Translated by R.A. Silverman and M.D. Friedman}, {Dover, New York}, pp.
  1-28.
\newblock {\em Uspekhi Matematicheskikj Nauk.}, 8:3--20, 1953.

\bibitem{abe2000axioms}
Sumiyoshi Abe.
\newblock Axioms and uniqueness theorem for tsallis entropy.
\newblock {\em Physics Letters A}, 271(1-2):74--79, 2000.

\bibitem{van1992stochastic}
Nicolaas~Godfried Van~Kampen.
\newblock {\em Stochastic processes in physics and chemistry}, volume~1.
\newblock Elsevier, 1992.

\bibitem{Tanimura2006Stochastic}
Yoshitaka Tanimura.
\newblock Stochastic {Liouville, Langevin, Fokker-Planck}, and master equation
  approaches to quantum dissipative systems.
\newblock {\em Journal of the Physical Society of Japan}, 75(8):082001, 2006.

\bibitem{Ge2016Mesoscopic}
Hao Ge and Hong Qian.
\newblock Mesoscopic kinetic basis of macroscopic chemical thermodynamics: A
  mathematical theory.
\newblock {\em Physical Review E}, 94(5-1):052150, 2016.

\bibitem{Jiang2005Mathematical}
Da~Quan Jiang, Min Qian, and Min~Ping Qian.
\newblock {\em Mathematical Theory of Nonequilibrium Steady-state}, volume
  1833.
\newblock Springer, New York, 2005.

\bibitem{oono1998}
Yoshitsugu Oono and Marco Paniconi.
\newblock Steady state thermodynamics.
\newblock {\em Progress of Theoretical Physics Supplement}, 130:29--44, 1998.

\bibitem{Esposito2010Three}
Massimiliano Esposito and Christian Van~den Broeck.
\newblock Three detailed fluctuation theorems.
\newblock {\em Physical Review Letters}, 104(9):090601, 2010.

\bibitem{Peng2018The}
Liangrong Peng, Yi~Zhu, and Liu Hong.
\newblock The {Markov} process admits a consistent steady-state thermodynamic
  formalism.
\newblock {\em Journal of Mathematical Physics}, 59(1):013302, 2018.

\bibitem{qian2007phosphorylation}
Hong Qian.
\newblock Phosphorylation energy hypothesis: open chemical systems and their
  biological functions.
\newblock {\em Annu. Rev. Phys. Chem.}, 58:113--142, 2007.

\end{thebibliography}
\end{document}